\documentclass[12pt]{iopart}
\usepackage{graphicx}
\usepackage{amssymb}

\begin{document}

\title[Measurement and Internalization of Systemic Risk in a Global Banking Network]{Measurement and Internalization of Systemic Risk in a Global Banking Network}

\author{Xiaobing Feng$^1$ and Haibo Hu$^2$}

\address{$^1$Shanghai University of International Business and Economics, Shanghai, 201620, P. R. China\\
fxb@sjtu.edu.cn
\\$^2$East China University of Science and Technology, Shanghai, 200237, P. R. China}
\begin{abstract}
The negative externalities from an individual bank failure to the whole system can be huge. One of the key purposes of bank regulation is to internalize the social costs of potential bank failures via capital charges. This study proposes a method to evaluate and allocate the systemic risk to different countries/regions using a SIR type of epidemic spreading model and the Shapley value in game theory. The paper also explores features of a constructed bank network using real globe-wide banking data.
\end{abstract}

\pacs{89.65.-s, 89.75.Hc, 87.23.Ge}

\section{Introduction}

There is no consensus on the definition of systemic risk in the literature. In this research the systemic risk is defined as contagion risk in the form of domino effect, which is caused by systemic rare events through systemically important institutions. The consequence of it is the huge negative externality on the economy and society. There are abundant studies on measurement of the systemic risk. In the review section of this paper, these literature are classified into two categories according to how the system is treated using different methodologies. The system studied is treated as a portfolio in the first category of the literature, hence ``the portfolio approach"; the system studied is treated as a complex system in the second category of the literature, hence ``the complex system approach". Research using the portfolio approach has experienced three stages of the development. In the earlier stage of the portfolio approach, contagion is defined as a shock in one country that is transmitted to another country after the common fundamental factors. Hence the residual correlation among countries after controlling the fundamentals is interpreted as contagion.$^{1-3}$ In the middle stage of the portfolio approach, Moreno \emph{et al.} treats equity as a call option based on a bank's assets.$^4$ Treating a bank's asset as options makes it possible to use large amount of time series trading data to measure contagion risk. Recently, using the idea of Shapley value,$^5$ suggests, without actual implementation, a way to measure and internalize the systemic risk.

There are however three problems in the current literature that need to be addressed and
can be improved: first the research has focused on the size of the banks (``the too big to fail case") rather than the linkages (``too connected to fail case"); second, the samples selected are only small part of the global banking system because of the data availability. The outcome based on the limited samples may not be reliable as a result of that; third, the common factors used to identify the systemic risk can hardly be exhausted, consequently the result may not be reliable either.

On the other hand, the research using the complex system approach has also experienced three stages of development. In the earlier stage, Allen and Gale claims that the possibility of contagion effect depends on the structure of interbank linkages.$^6$ They believe that a ``complete structure of linkages" will share the risk more easily than ``an incomplete structure" hence the risk sharing effect. Freixas \emph{et al.} further considers a structure of uni and multi-money center banking system, where the banks on the periphery are linked to the bank at the center but not to each other.$^7$ Multi-tiered banking system has then been examined with a similar approach. This research has shown that scholars have started to notice that the banking structure has impacts on contagion even though the models are simple and have not yet formally introduced the complex system theory. The second stage is symbolized by a conference entitled ``New Directions for Understanding Systemic Risk" which brought together experts from various disciplines to explore parallels between systemic risk in finance and systemic risk in engineering, ecology and other fields of sciences.  After that a series of interdisciplinary researches using various modeling and theory have emerged.$^{8-10}$ The 2007 US financial crisis becomes another driving force of this line of the research. At this stage researchers have started to apply complex system theory to disclose topology and features of different financial markets such as international trade network, investment network, and interbank clearing network.$^{10-14}$

There are again three issues in the current literature that need to be addressed and can be improved: first, the current research has focused on the disclosure of topology of the financial network, the studies on dynamics and the interaction between the dynamics and topology are limited; second, the application of knowledge from both economics and complex system is rather mechanical; third, the empirical research has mainly emphasized on two markets, interbank and payment systems, where data are relatively easy to obtain. Other markets, such as a global banking network, have hardly been researched, hence the focus of this research.

This paper intends to quantify the systemic risk of a global banking system using epidemic spreading model in complex system. Within this framework, the key instrument used to measure the systemic risk is Shapley value in game theory. The remaining parts of the research are arranged as follows: section two introduces the theoretical foundation of measurement and allocation of the systemic risk; section three illustrates the data description and the topology of the global banking network; section four presents the calculation results of the systemic risk and policy implication of global banking network governance; section five concludes the entire paper.

\section{The Measurement and Allocation of Systemic Risk}

\subsection{Quantifying the systemic risk of the global banking network using epidemic spreading model}

Disease spreading has been the subject of intense research since long ago.$^{15}$ In standard epidemic spreading modeling, population are classified into different compartments or states according to different spreading activities and different contacts. They are mainly ``susceptible", ``infected" and ``recovered". The modeling is based on two fundamental assumptions. First the homogeneous mixing hypothesis, which is equivalent to a mean-field treatment of the model. Second the time scale of the disease is much smaller than the lifespan of individuals, hence it is not necessary to include in the model terms accounting for the birth or natural death of agents.

The most widely used models in the epidemic spreading literature are ``Susceptible-Infected-Removable" (SIR) and ``Susceptible-Infected-Susceptible" (SIS) models. In these models based on a connected directed or undirected graph, the nodes represent agents that are in one of the states, and the edges represent the contacts between agents. Only susceptible agents in contact with one or more infected agents may become infected, infected agents can spontaneously become susceptible and then infected again in SIS model; infected agents can be recovered and never get infected again in SIR model.

In this research the agents are defined as nations' overseas banks and the domestic banks are not considered at this stage. Edges among agents represent the situation in which there is more than one overseas bank from country A to country B, or vise versa. ``Virus" is defined as ``default  Bank". Default is a state when banks are unable to observe their debt obligations and service their liabilities. The major reason of the default is the liquidity shortage within the bank. In the global banking network banks get infected by defaulted banks and claim default too. The defaulted banks will become financially sound again with internal or external rescue, basically in the form of capital injections. In practice, the infected banks are recovered with the aid of government funding. They can also be saved by merge \& acquisition from other financially sound banks. We assume that they will not get infected again within limited time period. Hence the SIR model is applied in this study as follows:
\begin{equation}
\left\{ \begin{array}{l}
 \frac{{{\rm{d}}I_k (t)}}{{{\rm{d}}t}} =  - \mu I_k (t) + \lambda kS_k (t)\Theta _k (\lambda ,t) \\
 \frac{{{\rm{d}}S_k (t)}}{{{\rm{d}}t}} =  - \lambda kS_k (t)\Theta _k (\lambda ,t) \\
 \frac{{{\rm{d}}R_k (t)}}{{{\rm{d}}t}} =  + \mu I_k (t) \\
 S_k (t) = 1 - I_k (t) - R_k (t) \\
 I_k (0) = I_k^0  \\
 \end{array} \right.
\label{1}
\end{equation}

These coupled equations can be interpreted as follows: defaulted
banks become the recovered ones at a rate of $ \mu$, while
non-defaulted banks become defaulted ones at a rate proportional to
both the densities of non-defaulted banks $S_k (t)$ and $\Theta _k
(\lambda ,t)$ which represents the probability that any given link
points to a defaulted bank. $\lambda$ is the microscopic spreading
rate, which is first assumed, then will be solved analytically and
represented as$^{4}$
\begin{equation}
\lambda _{\rm{c}}  = {{\left\langle k \right\rangle } \mathord{\left/
 {\vphantom {{\left\langle k \right\rangle } {\left\langle {k^2 } \right\rangle }}} \right.
 \kern-\nulldelimiterspace} {\left\langle {k^2 } \right\rangle }}
\label{2}
\end{equation}

The prediction of the model is the presence of a nonzero epidemic threshold $\lambda _{\rm{c}}$. If $\lambda > \lambda _{\rm{c}}$ the defaults spread and a financial crisis results. On the other hand if $\lambda < \lambda _{\rm{c}}$ the number of defaulted banks is infinitesimally small in the limit of a very large population, no financial crisis results. It shows that the $\lambda _{\rm{c}}$ is inversely proportional to the connectivity fluctuations. For regular networks, $\left\langle {k^2 } \right\rangle  < \infty $, the threshold has a finite value. On the other hand for networks with strongly fluctuating connectivity distributions, there is a vanishing epidemic threshold for increasing network size, i.e. $\left\langle {k^2 } \right\rangle  \to \infty $ for
$N \to \infty$.

The notion of thresholds forms a central part of classical and current epidemic theory and carries important implications for disease eradication and vaccination programs. Many studies demonstrate that social and spatial structures governing the connectivity of agents in networks have major control over the spread of infections and the emergence of intrinsic disease thresholds. In this study the epidemic threshold of the banking network is used to quantify the systemic risk of the global banking system, and to evaluate the systemic risk of individual banks in the network. The Shapley value will be adapted accordingly, which will be examined later in this paper.

\subsection{Quantifying and allocating the systemic risk of an individual bank in the network
using Shapley value}

While it is straightforward to measure the systemic risk of the whole system, it is less so when systemic risk assessment of individual banks in the network is considered.

The Shapley value (SV) methodology was developed in the context of
cooperative games in which the collective effort of a group of
players generates a shared ``value" (e.g. systemic risk, a negative
wealth) for the group as a whole.$^{16, 17}$ Given such a value, the
methodology decomposes it in order to allocate it across players
according to their individual ``contributions".

To apply SV to the banking network, it is necessary to specify a
``characteristic function $\Phi $" which maps each subsystem into a
systemic risk measure. This function is the same for all the subsets
of the system. The mapping in our case is the system of Eq. (1)
which maps the agents linked together as a network into a systemic
risk measure, i.e. the basic reproduction number. A bank network
that has $n$ banks will consist of $2^n$ subsystems which are:
$\varnothing$, \{1\}, \{2\}, $\cdots $, \{$n$\}, \{1, 2\}, \{1, 3\},
$ \cdots $, \{$n-1$, $n$\}, $ \cdots $, \{1, 2, $ \cdots $, $n$\}.
The ``contribution" of a bank ``$i$" to the entire banking system
equals the difference between the risk of subsystems excluding bank
``$i$" and the entire system including ``$i$". For a case of three
banks, the SV is defined as follows:
\begin{eqnarray}
{\rm{SV}}({\rm{bank}}_1 ) & = & (1/6)\{ 2 \times [\Phi \{ 1\}  - 0] + [\Phi \{ 2,\;1\}  - \Phi \{ 2\} ] + [\Phi \{ 3,\;1\}  - \Phi \{ 3\} ] \nonumber \\
& + & 2 \times [\Phi \{ 2,\;3,\;1\}  - \Phi \{ 2,\;3\} ]\}
\label{3}
\end{eqnarray}

A standard general formula for SV is as follows:
\begin{equation}
SV{}_i\left( \sum  \right) = 1/n\sum\limits_{n_s  = 1}^n {1/C\left( {n_s } \right)} \sum\limits_{\scriptstyle s \supset i \hfill \atop
  \scriptstyle |s| = n_s  \hfill} {\Phi (s) - \Phi \left( {s - \{ i\} } \right)}
\label{4}
\end{equation}
Where $\sum$ is the entire system, $s \supset i$ are all the subsystems in $\sum$ containing bank $i$, $|s|$ stands for the number of banks in subsystem $s$, and $C(n_s)$ is the number of subsystems comprising $n_s$ banks. In addition, the empty set carries no risk: $\Phi (\varnothing ) = 0$.

To compute the exact value of SV is an NP-Hard problem. In our case, there are 182 countries, there will be $2^{182}$ subsystems. It is essentially a complex combinatory issue. In this paper, we take the approximation which only considers the largest subsystem with 182 nations. This approximation simplifies the computation substantially.

\subsection{The challenge}

The idea of application of SV to evaluate the systemic risk of
individual economic agents was initiated by Ref. 16, but it has
never been implemented to any of the research based on either  the
portfolio approach or the complex system approach. Hence the focus
of this part of the research.

The challenge to apply SV method within the framework of the complex system lies in the fact that once a bank is left out of the banking system, the configurations of the remaining network vary tremendously.

First the remaining network will be split into a few subcomponents. Two rules can be used to calculate the SV of the subsystem. The SV value of the remaining system can be assessed based on either the average SV of the sub-components or the SV of the largest subcomponent. In this research the second approach has been applied and the corresponding algorithm to identify the largest subcomponents is used accordingly.

We need to compute $2^{182}-1$ terms which  is practically not feasible in terms of computational complexity. Thus it is hard to obtain the exact SV for each country/region. However $\lambda _{\rm{c}} $ is still an appropriate metric quantifying the importance of each country/region.
Thus $\lambda _{\rm{c}} $ of the largest subcomponents will be calculated.

Second, it is noted that the topology of the largest subcomponent of remaining network may or may not be the same as that of the original network. In particular, the degree distribution can be quite different in some cases. In this research, the Kolmogorov-Smirnov (K-S) test are proposed to be implemented to examine whether the two samples are drawn from the same distributions.
We find that for all 182 countries/regions K-S test gives $H=0$, and actually $p$ is equal to or approximately equal to 1 at a significance level of 0.01. Therefore the nature of the network has not changed after the deletion of a node, we can use the same formula for the calculation of the systemic risk.

\section{Topology of the Global Banking Network}
\subsection{The database}

The data used in this research is from Bankscope which has information on over 30,000 public and private banks throughout the world from 2009 to 2011. Each bank report contains detailed consolidated and unconsolidated balance sheets and income statements. Data comes from Fitch Ratings and six other sources. It also provides company and country risk ratings and reports, ownership, and security and price information. This database is produced by Bureau van Dijk.

The information in the database that can be used for network construction are the number of branches and subsidiaries that each parent bank has established overseas. Subsidiaries are banks that are completely or partly owned and wholly controlled by parent banks that own more than half of the subsidiary's stock. For the purpose of this research subsidiary bank and branch banks are equally treated as overseas banks. The information on the overseas location of the two types of banks can be found in ``ownership" category of the database.

If bank ``$i$" in country A has $n$ banks in country B, it indicates that bank ``$i$" in country A provides or exports service ``products" to country B, therefore $n \ge 1$ links are established between country A and B. If there are $m$ banks in country A that do the same as bank ``$i$" in country A, then there will be $m \times n$ links between country A and B. On the other hand, country B can also set up branches or subsidiaries in A, which indicates that A imports bank-service from B. It is to be noticed that only the directions are different in these two cases. We did the same for 182 countries/regions and a directed and weighted global bank network based on edges is therefore established.

The database also provides the balance sheets information of each bank within which there are two accounting items named ``Loans and advances to banks" and ``Deposit from banks". They are records of when banks trade in the interbank markets. The position varies in a high frequency. However it is almost impossible to use them for network construction for two reasons: first, it is highly unlikely to identify the proportional position of a bank to its specific counterparties. There is only a lump sum trading position e.g. ``Loans and advances to banks" of HSBC Holding Company. Second, even if such a kind of specific data is available from certain super-regulators who monitor all transaction and have access to the information, they might be meaningless because the market-maker in interbank market ``pools" the surplus funding from all banks and then directs them to the needed ones, i.e. there are no-explicit intended one to one counterparties specification. Hence in this research we will use edge based network instead of interbank loan based network for network construction.

\subsection{The topology of the global banking network}

The network in Figure 1 is drawn using Gephi. The network is almost fully connected with 6 islands. The top five countries in terms of vertex total strengths in global network are US, UK, France, Germany and Switzerland in that order. The top five countries/regions in terms of vertex strengths in Asia are China, Hong Kong, Japan, Singapore and Bahrain in that order. In terms of in-degree the top five countries are UK, US, Russia, Switzerland and France. It is interesting to note that Panama Island is ranked sixth as one of the famous offshore banks. The top five countries/regions in Asia are Indonesia, Singapore, Israel, Hong Kong and Malaysia. These are the export deficit countries in exporting banking service trades. In terms of out-degree the top five countries are US, UK, France, Germany and the Netherlands. In Asia they are Japan, China, India, Bahrain and Kuwait. As historical Asian international financial centers, Singapore, Japan and Hong Kong are still playing important roles, and China is gaining its momentum. These are the export surplus countries in exporting banking service trades. It shows that the rankings are different when different parameters are considered. In particular, the network shows that the in-degree number of a country/region is different from its out-degree number. It implies that the export of banking services of a nation is not equal to the imports of banking service of the country. Hence trade imbalances are observed in the global banking network.

\begin{figure}
  \centerline{\includegraphics[height=3.5in]{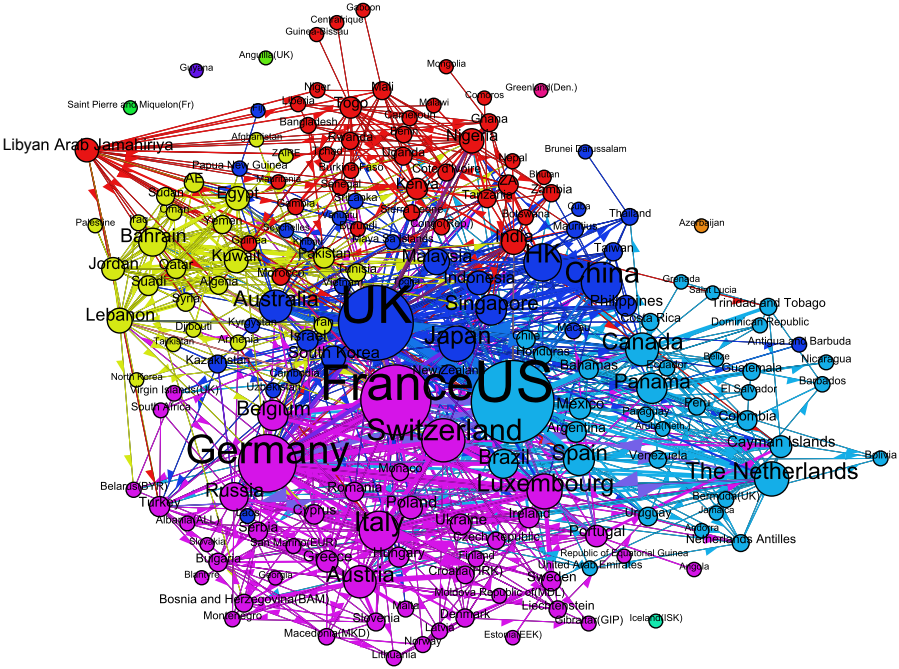}}
  \caption{The weighted global banking network. The width of edge
indicates its weight and color is the mixing of its source node color and target node color. The size of node indicates its strength and color represents its modularity
class. \label{f1}}
\end{figure}

In Figure 1 by denoting different colors we then identify 11
clusters using the community detection method developed in Ref. 18.
It seems that the clusters are geographically related. This finding
does not support the argument in the literature that distances do
not play roles with the development of electronic banking. Table 1
in the appendix lists the country/region names within each community. It
indicates that community 10 is mainly a European based bank block;
community 3 is mainly an America based bank block; community 4 is an
Asian Pacific based bank block; community 5 is an oil produce OPEC
bank block; community 7 is an African bank block respectively.

\section{Calculation of the Systemic Risk and Internalization}
\subsection{Systemic risk of each country/region}

In this section, we calculate the systemic risk for each country/region based on the theoretical model in section 2 and the global banking network constructed in section 3.

Figure 2 presents the systemic risk that each country/region generates and contributes to the entire global banking system. The corresponding values are listed in Table 2 in the appendix. As a benchmark, the horizontal line cutting in the middle in Figure 2 is the systemic risk of the original global network without deleting any country/region. The vertical lines are systemic risk measured by $\lambda _{\rm{c}} $ of each country/region. It shows that the majority of the countries/regions have made ``contributions" to the systemic risk of the system since they are above the benchmark line. There are however a few countries/regions, that are below the cutting line, which have made ``negative contributions" to the systemic risk of the system. The top five contributors to global systemic risk are the UK, US, France, Germany and Switzerland, which is consistent with the rank of degree. They are considered as systemically important institutions in the system. However the relationship between systemic risk and the degree of each country/region may not be linear, see appendix for the plot of systemic risk and degree.

\begin{figure}
  \centerline{\includegraphics[height=3in]{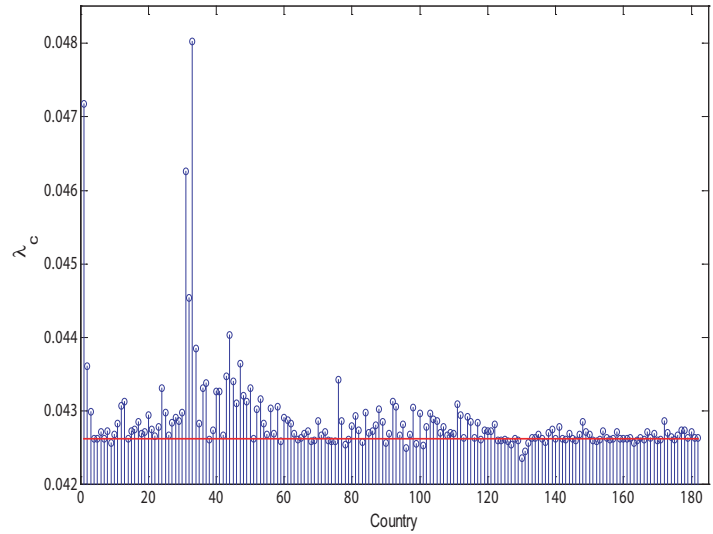}}
  \caption{The country/region-specific systemic risk. \label{f2}}
\end{figure}

\subsection{Internalization of the systemic risk}
The current approach to bank regulation implicitly assumes that the system as a whole can be made safe if the safety of the individual bank is ensured. The capital requirement ratio set by the Bank of International Settlements is targeted to individual banks. The lesson from the most recent US financial crisis however is that in trying to make individual banks themselves safer, banks, can behave in a way which collectively undermines the whole system.

The systemic risk causes huge negative externalities and social cost. We believe that one of the key purposes of bank regulation is to internalize the social cost of potential bank failures via capital adequacy requirements or capital charges. Countries that ``contribute" to the systemic risk should be responsible for it by paying a certain amount of capitals. In this section we will specify the amount of capital that is required for compensating systemic risk generated. In this way, the safety of the entire banking system can be ensured.

In section 4.1 the country level systemic risk has been calculated. In this section the evaluated systemic risk will be converted into country level capital charge for compensation by using a type two Holling function. The converted capital charge is defined as effective capital charge.

Holling functions are functional response in ecology. They are a relationship between the intake rate of a consumer and food density. Type two Holling function is characterized by a decelerating intake rate, which follows from the assumption that the consumer is limited by its capacity to process food. When the density of prey population is small, the number of prey consumed increases fast. The speed of the rate change increases then decreases as the density increases in order to maintain the stability of the system.

\begin{figure}
  \centerline{\includegraphics[height=2in]{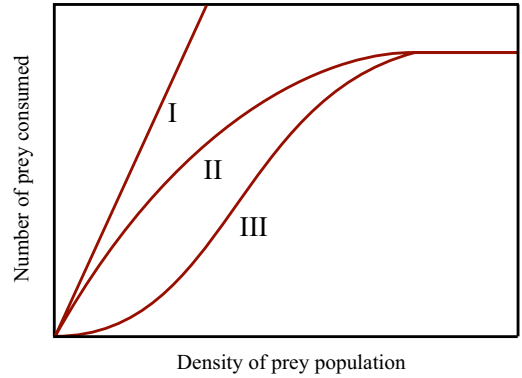}}
  \caption{Shapes of type I, II and III Holling functions. Note: The
density of prey population is ranked in ascending order. \label{f3}}
\end{figure}

In the banking network, type two Holling function (Figure 3) indicates when a country has higher $\lambda _{\rm{c}} $, i.e. in general a higher density of bank linkages for a country (the so called prey), it will impose higher systemic risk onto the network as a result. It hence requires a ``machine" ( the so called predator dictated by the banking system regulator) inside the system to absorb (consume) the systemic risk in order to maintain the sustainability of the system. But the effectiveness of the machine to absorb the systemic risk is limited, therefore the systemic risk consumed by the machine increases at a descending order. The formula used to calculate the capital charges is presented in
\begin{equation}
{\rm{Capital\,Charge  = }}{{\lambda _{\rm{c}}^n } \mathord{\left/
 {\vphantom {{\lambda _{\rm{c}}^n } {\left( {1 + \lambda _{\rm{c}}^n } \right)}}} \right.
 \kern-\nulldelimiterspace} {\left( {1 + \lambda _{\rm{c}}^n } \right)}} \label{4}
\end{equation}
where $n=1, 2, \cdots, N $.

The results calculated based on Eq. (5) are presented in Figures (4) and (5). It shows that such countries as the UK and the US with higher bank density are required to be charged with higher capitals as they are considered as systemically important institutions. It can be shown from the right hand side of the tips of the curves.

The implication here is that the capacity for a system to absorb systemic risk is not infinite, the systemically important countries may not be able to expand their overseas banking sub-networks based solely on their own objective functions, and without any constraints from the perspective of the global network stability. The capital charge designed in this paper is one instrument to regulate the systemic stability.

\begin{figure}
  \centerline{\includegraphics[height=2in]{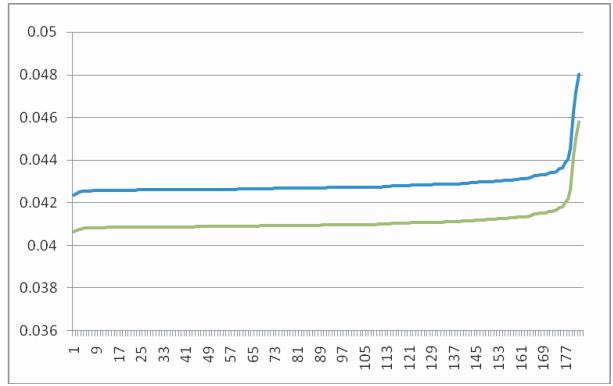}}
  \caption{The systemic risk and the effective capital charge. Note:
The horizontal axis is the countries/regions ranked according to the
systemic risks. The top line is the $\lambda _{\rm{c}} $ that is
related to the number of banks linked to a country/region, and the
bottom line is the converted effective capital charges for each
country/region. \label{f4}}
\end{figure}

\begin{figure}
  \centerline{\includegraphics[height=2in]{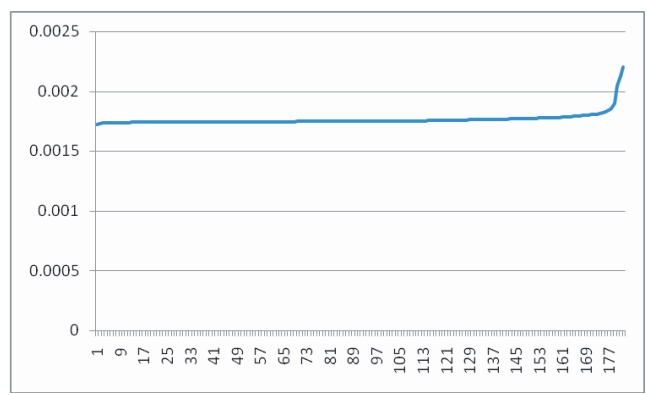}}
  \caption{The difference between systemic risk and effective capital
charge. Note: the difference line has a tip on the right hand side
of the curve, which indicates that higher capital charges are
required for systemically important institutions. \label{f5}}
\end{figure}

\section{Conclusion and Discussion}
Based on SIR-type-of-epidemic spread modeling within the framework of complex system theory, using the Shapley value in cooperative game theory, this research evaluates the systemic risk of a constructed global banking network with real data.

The major findings are that the distribution of systemic risk among countries/regions varies dramatically, but five countries UK, US, France, Germany and Switzerland account for the majority of the risk at the global level and Japan, China, Singapore, India and South Korea account for the majority of the risk in Asia. The magnitude of the systemic risk at the national level is related to the degree distribution of banks in a nonlinear fashion. To be more specific, it depends on whether the network is more heterogeneous such as a scale-free network, or more homogeneous such as an exponential or even a regular network.

The constructed global banking network includes over 30, 000 public and private overseas banks all over the world. The systemically important institutions are identified. The detected modularity of the global network indicates that the geographical location still plays a role in formulating the communities.

The systemic risk is internalized by capital charges required from each country/region. The capital charge is evaluated based on the country/region level systemic risk. A type-two-Holling function is used to convert systemic risk to capital charge.

Finally we suggest that individual risk control policy should be combined with the systemic risk control policy to maintain the stability of the system, neither of which can be ignored. This is advice that is different from the current policy stance that emphasizes only the safety of individual banks.


\section*{Appendices}

\begin{table}[t]
\caption{Countries/regions within different communities (Only sample countries/regions are provided, details are available from the author upon request).}
\center
{\begin{tabular}{@{}llc@{}}
\br
Code & Country/Region  & Modularity class \\
\mr
31 & France         &                               10 \\
32 & Germany         &                              10 \\
35 & Portugal        &                              10 \\
42 & Sweden          &                                 10 \\
43 & Belgium          &                                 10 \\
\mr
164 & Liberia                &                            7 \\
166 & Malawi                &                            7 \\
167 & Mauritania           &                              7 \\
146 & Comoros             &                              7 \\
147 & Congo (Rep.)         &                               7 \\
\mr
115 & Jordan              &                               5 \\
116 & Qatar                &                              5 \\
64 & Saudi            &                                   5 \\
70 & Kuwait               &                               5 \\
71 & Iran                 &                               5 \\
\mr
45 & China           &                                    4 \\
46 & HK                  &                               4 \\
47 & Japan         &                                      4 \\
61 & Taiwan        &                                      4 \\
63 & Vietnam         &                                    4 \\
\mr
1 & US                                &                   3 \\
2 & Canada          &                                     3 \\
3 & Mexico       &                                        3 \\
23 & Colombia       &                                     3 \\
24 & Brazil            &                                    3 \\
\br
\end{tabular} \label{app1}}
\end{table}

\begin{table}[t]
\caption{Systemic risk measured by $\lambda _{\rm{c}} $ by countries
(Only a sample is provided, details are available from the author
upon request).} 
\center
{\begin{tabular}{@{}lll@{}} 
\br
Code & Country  & $\lambda _{\rm{c}} $ \\
\mr
33 & UK &   0.048019 \\
1 & US &    0.047179 \\
31 & France &   0.04626 \\
32 & Germany &  0.044538 \\
44 & Switzerland &  0.044037 \\
34 & The Netherlands &  0.043847 \\
47 & Japan &    0.043644 \\
2 & Canada &    0.043616 \\
43 & Belgium &  0.043468 \\
76 & Australia &    0.043422 \\
45 & China &    0.043408 \\
37 & Italy &    0.043384 \\
50 & Singapore &    0.043315 \\
24 & Brazil &   0.043313 \\
36 & Russia &   0.043308 \\
40 & Spain &    0.043266 \\
41 & Luxembourg  &  0.043266 \\
48 & India  &   0.043211 \\
53 & South Korea &  0.043159 \\
49 & Indonesia &    0.043135 \\
92 & Poland  &  0.043133 \\
13 & Panama &   0.04313 \\
\br
\end{tabular} \label{app2}}
\end{table}

\begin{figure}
  \centerline{\includegraphics[height=2.5in]{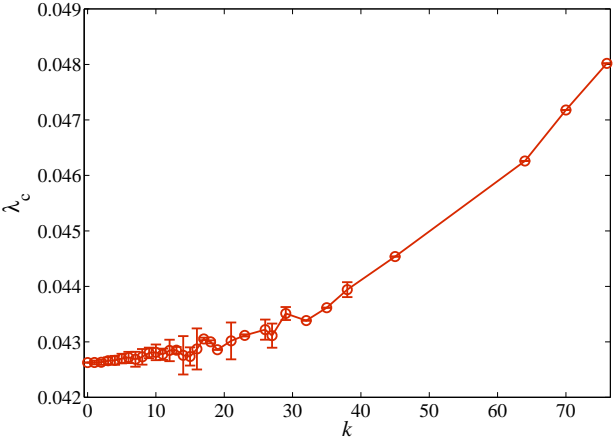}}
  \caption{The relation between systemic risk $\lambda _{\rm{c}} $ and
degree for 182 countries/regions. Error bars with $ \pm 1$
standard deviation are shown. It is obvious that the relation is not
linear, and systemic risk almost becomes a flat line for small degree. \label{app3}}
\end{figure}


\end{document}